\documentclass[epj]{svjour}
\usepackage{graphics}
\usepackage{amsfonts}
\usepackage{amssymb}
\usepackage{amsmath}
\usepackage{graphicx}
\usepackage{float}
\usepackage{caption}
\usepackage{multicol}
\usepackage{dcolumn}

\begin{document}

\title{Gravitational Waves from Preheating in Gauss-Bonnet Inflation}
\author{K. El Bourakadi\inst{1,2}, M. Ferricha-Alami\inst{1,2}, H. Filali%
\inst{2}, Z. Sakhi\inst{1,2} \and M. Bennai\inst{1,2}\thanks{\emph{Present
address:} k.elbourakadi@yahoo.com}}
\offprints{}
\institute{Physics and Quantum Technology Team, LPMC, Ben M'sik Faculty of Sciences, 
{\small Casablanca Hassan II University, Morocco \and LPHE-MS Laboratory
department of Physics, Faculty of Science, Mohammed V University in Rabat,
Morocco}}
\date{Received: date / Revised version: date}

%

%

%
\abstract{
We study gravitational wave production in an expanding Universe during the
first stages following inflation, and investigate the consequences of
the Gauss-Bonnet term on the inflationary parameters for a power-law\ inflation
model with a GB coupling term. Moreover, we perform the analyses on the
preheating parameters involving the number of e-folds $N_{pre}$, and the
temperature of thermalization $T_{th},$ and show that it's sensitive to the
parameters $n$,\ and $\gamma$, the parameter $\gamma$ is proposed to connect the density
energy at the end of inflation to the preheating energy density.\ We set a
correlation of gravitational wave energy density spectrum with the spectral
index $n_{s}$ detected by the\ cosmic microwave background experiments$.$
The density spectrum $\Omega _{gw}$ shows good consistency with observation
for $\gamma =$\ $10^{3}$\ and $10^{6}$. Our findings suggest that the
generation of gravitational waves (GWs) during preheating can satisfy the
constraints from Planck's data.
\PACS{
      {PACS-key}{discribing text of that key}   \and
      {PACS-key}{discribing text of that key}
     } } 
\maketitle
\section{Introduction}

\label{intro} In the very early stages of the Universe's evolution,
inflation is the leading paradigm that was proposed to resolve issues namely
flatness and horizon problems that appear in the standard big bang
cosmological model. During inflation, tensor modes are produced from the
amplification of initial quantum fluctuations into classical perturbations
outside the Hubble radius, due to the accelerated expansion of the universe 
\cite{G1}. They may cause a $B$-mode polarization of the cosmic microwave
background $CMB$ photons. As a result, observations $CMB$ can be used to
constrain the amplitude of the tensor perturbations and inflationary models
can be strongly constrained using the combination of $n_{s}$ and $r$. In the
first stage following inflation, preheating is characterized in most models
by an explosive and non-perturbative generation of non-thermal fluctuations
of the inflaton and other bosonic fields connected to it \cite{G3}. In
chaotic models of\ inflation, the inflaton decays via parametric resonant
particle creation \cite{G4}, accompanied by violent dynamics of non-linear
inhomogeneous structures of the scalar fields \cite{G5}. Preheating can
accelerate the thermalization of our universe since the inflaton energy can
be transferred rapidly into radiation matter. Thus, This period of a rapid
particle production is highly inhomogeneous and generically generates
gravitational waves with large energy densities \cite{G7,G8}. As a
consequence, the detection of GWs generated during preheating can help us to
test inflation and understand the process of reheating. According to general
relativity, the current universe should be penetrated by a diffuse
gravitational wave background (GWB) coming from several sources like relic
stochastic backgrounds from the early universe, phase transitions,
inflation, turbulent plasmas, cosmic strings, etc. \cite{G10}. These
backgrounds have very different spectral shapes and amplitudes that may, in
the future, allow gravitational wave observatories like LIGO, LISA, BBO, or
DECIGO \cite{G10} to disentangle their origin. Cosmological gravitational
wave background could potentially carry original and pure information about
the universe at early times. For low energy scale inflationary models, the
frequencies of the gravity waves generated after inflation may occur in the
range that can be detected in theory by direct detection tests, providing us
with a channel for verifying inflation from\ the $CMB$ data.

Another extended theory of inflation that has been studied is a scalar field
coupled to the Gauss-Bonnet combination of quadratic curvature scalars $%
R_{GB}^{2}$ \cite{G11,G12}. In this paper, we study whether a family of such
theories, specifically the Gauss-Bonnet theory coupled with functions of a
scalar field, may accurately predict inflationary dynamics compatible with
current observational constraints on the parameters of these theories. It
was also claimed that the temperature of reheating and the equation-of-state
(EoS) parameter during reheating can be probed by looking at the spectrum of
the GW background \cite{G13,G14}. Therefore, in this work, we consider
inflationary models with a Gauss-Bonnet (GB) term to estimate the energy
spectrum of the PGW and to provide constraints on the preheating parameters.

The paper is organized as follows. In section II, we develop the basic
equations that describe a GB inflation model. In section III, we calculate
the expression for the energy spectrum of these GWs. We further perform
constraints on the preheating parameters in section IV. In section V, we
converte the spectra into physical variables and describe gravitational
waves from Planck's measurements point of view. We conclude in section VI.

\section{The Gauss-Bonnet Model}

\label{sec:1} We consider the following action that involves the
Einstein-Hilbert term and the GB term coupled to a canonical scalar field $%
\phi $ through the coupling function $\xi (\phi )$\ \cite{Ga1,G11},

\begin{eqnarray}
S &=&\int d^{4}x\sqrt{-g}\left[ \frac{1}{2\kappa ^{2}}R-\frac{1}{2}g^{\mu
\nu }\partial _{\mu }\phi \partial _{\nu }\phi \right]  \notag \\
&&-\int d^{4}x\sqrt{-g}\left[ V(\phi )+\frac{1}{2}\xi (\phi )R_{GB}^{2}%
\right]
\end{eqnarray}%
where $R_{GB}^{2}=R^{2}-4R_{\mu \nu }R^{\mu \nu }+R_{\mu \nu \rho \sigma
}R^{\mu \nu \rho \sigma }$ is the GB term and$\ \kappa ^{2}=8\pi
G=M_{p}^{-2}.$ The model is hence specified by two arbitrary functions, the
potential $V(\phi )$ and the Gauss-Bonnet coupling $\xi (\phi )$. The
background dynamical equations for inflation with the GB term which couples
to a scalar field $\phi $ in a spatially flat FRW Universe are

\begin{equation}
\frac{3H^{2}}{\kappa ^{2}}=\frac{1}{2}\dot{\phi}^{2}+V(\phi )+12\dot{\xi}%
H^{3},  \label{1}
\end{equation}

\begin{equation}
-\frac{2\dot{H}}{\kappa ^{2}}=\dot{\phi}^{2}-4\ddot{\xi}H^{2}-4\dot{\xi}H(2%
\dot{H}-H^{2}),  \label{2}
\end{equation}

\begin{equation}
\ddot{\phi}+3H\dot{\phi}+V_{,\phi }+12\xi _{,\phi }H^{2}(\dot{H}+H^{2})=0,
\label{3}
\end{equation}%
the dot represents a derivative with respect to the cosmic time $t$, $H=\dot{%
a}/a$ denotes the Hubble parameter, and $V_{,\phi }=\partial V/\partial \phi
,$\ $\xi _{,\phi }=\partial \xi /\partial \phi ,$\ $\xi $\ is a function on$%
\ \phi ,\ $and $\dot{\xi}=$\ $\xi _{,\phi }\dot{\phi}.$

\bigskip The so-called slow-roll parameters are expressed in terms of the
potential and the coupling functions as

\begin{eqnarray}
\epsilon &\approx &\frac{Q}{2}\frac{V_{,\phi }}{V},  \label{a} \\
\eta &\approx &-Q\left( \frac{V_{,\phi \phi }}{V_{,\phi }}-\frac{V_{,\phi }}{%
V}+\frac{Q_{,\phi }}{Q}\right) ,  \label{b} \\
\delta _{1} &\approx &-\frac{4}{3}\xi _{,\phi }QV,  \label{c} \\
\delta _{2} &\approx &-Q\left( \frac{\xi _{,\phi \phi }}{\xi _{,\phi }}+%
\frac{V_{,\phi }}{V}+\frac{Q_{,\phi }}{Q}\right) ,  \label{d}
\end{eqnarray}%
with $Q\equiv V_{,\phi }/V+(4/3)\xi _{,\phi }V.$ The potential and the
coupling function can also be used to define the e-folding number $N$ at the
horizon exit before the completion of inflation,

\begin{equation}
N_{k}\approx \int_{\phi _{end}}^{\phi _{k}}\frac{3V}{3V_{,\phi }+4\xi
_{,\phi }V^{2}}\kappa ^{2}d\phi =\int_{\phi _{end}}^{\phi _{k}}\frac{\kappa
^{2}}{Q}d\phi ,  \label{N}
\end{equation}%
where the subscripts \textquotedblleft $k$\textquotedblright\ and
\textquotedblleft $end$\textquotedblright\ respectively indicate the moment
when a mode $k$ crosses the horizon and the end of inflation.\bigskip

The spectral indices of scalar and tensor perturbations $n_{s},$ and
tensor-to-scalar ratio $r$ are calculated as \cite{Ga3}

\begin{eqnarray}
n_{s}-1 &\simeq &-2\epsilon -\frac{2\epsilon (2\epsilon +\eta )-\delta
_{1}(\delta _{2}-\epsilon )}{2\epsilon -\delta _{1}},  \label{4} \\
r &\simeq &8(2\epsilon -\delta _{1}).  \label{6}
\end{eqnarray}%
Choosing the form of the potential $V(\phi )$ and the coupling function $\xi
(\phi )$ and using Eqs. (\ref{a}-\ref{d}), (\ref{4}-\ref{6}), the
theoretical predictions of any particular inflation model can be verified
using observational data \cite{G2}.

\subsubsection*{power-law model with inverse monomial coupling}

Let us consider a power-law model of GB inflation with inverse monomial
coupling. The inflaton potential and the coupling function are given by

\begin{equation}
V(\phi )=V_{0}(\kappa \phi )^{n},~~~\xi (\phi )=\xi _{0}(\kappa \phi )^{-n},
\end{equation}%
here, $n$ is assumed to be positive, and $V_{0},\xi _{0}$\ are a
dimensionless constants. This model has received a lot of attention \cite%
{G11,Ga3}, were they establish an analytic relationship between the spectral
index of curvature perturbations and the tensor-to-scalar ratio thanks to
the specific choice of GB coupling. From Eqs. (\ref{a})-(\ref{d}) and (\ref%
{N}), the observable quantities in Eqs. (\ref{4}-\ref{6}) can be obtained in
terms of $N_{k}$ as

\begin{equation}
n_{s}-1=-\frac{2(n+2)}{4N_{k}+n},~~r=\frac{16n(1-\alpha )}{4N_{k}+n},
\end{equation}%
where $\alpha \equiv 4V_{0}\xi _{0}/3$. For that case, we conclude that such
a specific choice of GB coupling allows us to find an analytic relation
between $r$, $n_{s}$ in terms of $N_{k}.$

\begin{figure*}[h]
\resizebox{1\textwidth}{!}{  \includegraphics{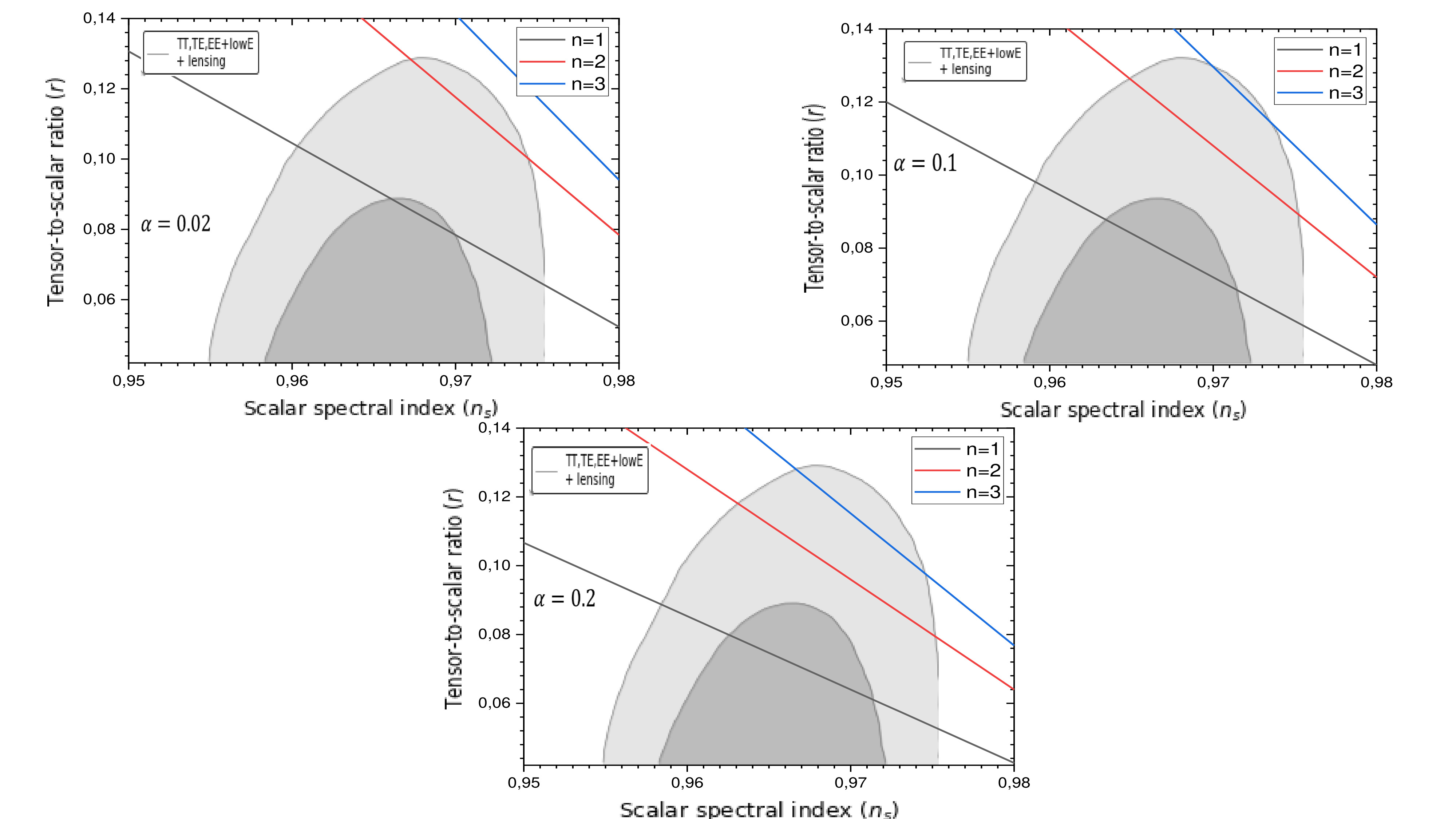}
}
\caption{$r$ versus $n_{s}$ plot for monomial inflation with a Gauss-Bonnet
coupling function. Inner and outer shaded regions are $1\protect\sigma $ and 
$2\protect\sigma $ constraints from Planck respectively. We choose three
different values of $n$, for the black line $n=1$, the red line represents $%
n=2$ and for the blue line, $n=3$.}
\label{fig:1}
\end{figure*}

In Fig. \ref{fig:1}, we consider the usual power-law inflation, where the
inflaton field is coupled with the Gauss-Bonnet term through the inverse
monomial model. It is apparent from these plots that increasing the value of 
$\alpha $\ to $0.2$ makes the decreasing function obtain from our model in
certain regions consistent with the latest observations in certain regions
for $n=1,2,\ $and $3.$

\section{Primordial Gravitational Waves}

\label{sec:2} The intense production of matter fields after inflation can
promote substantial metric changes. However, we are only interested in the
evolution of the transverse-traceless satisfied be the metric perturbation $%
h_{ij}$, and the generation of GWs during preheating. Therefore, GWs can be
represented by the traceless part of the spatial metric perturbations in the
FRW background \cite{Ga4,Ga5}:

\begin{equation}
ds^{2}=g_{ij}dx^{i}dx^{j}=-dt^{2}+a(t)^{2}\left( \delta _{ij}+h_{ij}\right)
dx^{i}dx^{j}.
\end{equation}%
The perturbation $h_{ij}$ satisfies the transverse-traceless (TT)
conditions: $\partial _{i}h_{ij}=h_{ii}=0,$\ and has the equation of motion

\begin{equation}
\ddot{h}_{ij}+3H\dot{h}_{ij}-\frac{1}{a^{2}}\nabla ^{2}h_{ij}=2\kappa
^{2}S_{ij}^{TT},  \label{7}
\end{equation}%
the source term $S_{ij}^{TT}$\ is the transverse-traceless of the
anisotropic stress $S_{ij}$.

\subsubsection*{Gravitational Wave Energy Density}

The energy density power spectrum of GWs sourced by the inhomogeneous decay
of the symmetry braking field, can be defined as the energy density averaged
over a volume $V$ of several wavelengths size \cite{G3}, This energy density
carried by GWs can be calculated through the following equation \cite{Ga6}

\begin{equation}
\rho _{gw}=\frac{1}{4\kappa ^{2}}\left\langle \dot{h}_{ij}(t,\mathbf{x})\dot{%
h}_{ij}(t,\mathbf{x})\right\rangle .
\end{equation}

The strength of GW is characterized by their energy spectrum, which
represents the abundance of gravity wave energy density today, is given as

\begin{equation}
h^{2}\left( \frac{\rho _{gw,0}}{\rho _{c,0}}\right) =\int \frac{df}{f}%
h^{2}\Omega _{gw,0}(f),
\end{equation}%
which can be rewritten as

\begin{equation}
h^{2}\Omega _{gw,0}(f)=\frac{h^{2}}{\rho _{c,0}}\frac{d\rho _{gw,0}}{d\ln f},
\end{equation}%
where $f$ is the frequency and $\rho _{c,0}=3H_{0}^{2}/(8\pi G)$\ is the
critical energy density today.

Next, we need to consider the evolution of the scale factor during
preheating that is parameterized by an $e$-folds number $N_{pre}$\ and test
its dependency on the evolution of the equation of state. In general, when
the inflaton field oscillates around its minimum, the equation of state
jumps from $\omega =0$ to an intermediate value close to $\omega =1/3$
during preheating \cite{Ga7,Ga8}.

\section{Preheating Constraints}

\label{sec:3} The process of preheating happens in the early stages of the
Universe's evolution. This is thought to be necessary because the universe
cools as it expands. As a result, there must be a period immediately
following inflation to allow it to thermally prepare for the next step,
which we call preheating. Preheating occurs due to the interaction of
massless scalar field $\chi $ with the oscillating inflaton field, which
causes it to grow exponentially fast, as a result of parametric resonance,
followed by a stage of thermal equilibrium (reheating) \cite{Ga9}. To
extract information about preheating we need to consider the phase between
the time observable CMB scales crossed the horizon and the present time.
Deferents eras occurred throughout this length of time which can be
described by the following equation :

\begin{equation}
\frac{k}{a_{0}H_{0}}=\frac{a_{k}}{a_{end}}\frac{a_{end}}{a_{pre}}\frac{%
a_{pre}}{a_{th}}\frac{a_{th}}{a_{eq}}\frac{a_{eq}}{a_{0}}\frac{H_{eq}}{H_{0}}%
\frac{H_{k}}{H_{eq}},  \label{9}
\end{equation}%
here, $a_{0},a_{k},a_{end},a_{pre},a_{th}$ end $a_{eq}$ respectively
correspond to the scale factor at present, time of horizon crossing, end of
inflation, end of preheating, and the time of thermal equilibrium, and
finally, the end of the matter and radiation equality era, whereas $H_{0}$
and $H_{eq}$ are the Hubble constant at present time and the time of matter
and radiation equality. Taking the number of e-folds $N$ into consideration,
we can rewrite Eq. (\ref{9}) as 
\begin{equation}
\ln \left( \frac{k}{a_{0}H_{0}}\right) =-N_{k}-N_{pre}-N_{th}+\ln \frac{%
a_{th}}{a_{0}}+\ln \frac{H_{k}}{H_{0}},  \label{e}
\end{equation}%
the number of e-folds between the time when a mode exits the horizon and the
end of inflation is parametrized by $N_{k}=a_{end}/a_{k},$ and $%
N_{pre}=a_{pre}/a_{end}$ is the duration from the end of inflation to the
end of preheating, finally, $N_{th}=a_{th}/a_{pre}$ is the number of e-folds
between the end of preheating and the thermal equilibrium (end of
reheating). Our goal is to calculate the preheating duration $N_{pre}$\ in
terms of inflationary parameters. Considering that no entropy production
occurred after the thermal equilibrium was completed, one can write \cite%
{Gg2}

\begin{equation}
\frac{a_{th}}{a_{0}}=\frac{T_{0}}{T_{th}}\left( \frac{43}{11g_{\ast }}%
\right) ^{\frac{1}{3}},  \label{f}
\end{equation}%
where $T_{0}$\ is the current temperature of the Universe, and $T_{th}$\ is
the thermal equilibrium temperature of reheating, the energy density $\rho
_{th}$ at the end of reheating is defined as :

\begin{equation}
\rho _{th}=\frac{\pi ^{2}}{30}g_{\ast }T_{th}^{4},  \label{g}
\end{equation}%
where $g_{\ast }$ is the number of relativistic degrees of freedom at the
end of reheating. Using the expressions $\rho _{th}\propto
a_{th}^{-3(1+\omega )}$\ and $\rho _{end}\propto a_{end}^{-3(1+\omega )}$\
that respectively corresponds to the reheating and inflation energy
densities. We assume that the energy density at the end of inflation and
preheating energy density are related by a parameter $\gamma $

\begin{equation}
\rho _{end}=\gamma \rho _{pre}=\gamma ~a_{pre}^{-3(1+\omega )},
\end{equation}%
then,

\begin{equation}
\frac{\rho _{end}}{\rho _{th}}=\gamma \left( \frac{a_{pre}}{a_{th}}\right)
^{-3(1+\omega )},
\end{equation}%
writing this in terms of e-foldings, one can obtain

\begin{equation}
\rho _{th}=\frac{\rho _{end}}{\gamma }~e^{-3(1+\omega )~N_{th}},  \label{h}
\end{equation}%
the energy density of inflation $\rho _{end}$ is determined by the potential
at the end of inflation $V_{end}$ and $\lambda _{end}$ given as follows :

\begin{equation}
\rho _{end}=\lambda _{end}V_{end},
\end{equation}%
the effective ratio of kinetic energy to potential energy $\lambda _{end}$\
is calculated from the GB\ field equation (\ref{3}) \cite{Gb5}:

\begin{equation}
\lambda _{end}=\left( \frac{6}{6-2\epsilon -\delta _{1}\left( 5-2\epsilon
+\delta _{2}\right) }\right) _{\phi =\phi _{end}}.  \label{j}
\end{equation}

We can derive a total duration using Eqs. (\ref{e}), (\ref{f}), (\ref{g}),
and (\ref{h}) :

\begin{eqnarray}
N_{pre}+\frac{1-3\omega }{4}N_{th} &=&-\ln \left( \frac{k}{a_{0}T_{0}}%
\right) -\frac{1}{3}\ln \left( \frac{11\bar{g}_{\ast }}{43}\right)  \label{i}
\\
&&-\frac{1}{4}\ln \left( \frac{30\lambda _{end}}{\gamma \pi ^{2}\bar{g}%
_{\ast }}\right) -\frac{1}{4}\ln \left( \frac{V_{end}}{H_{k}^{4}}\right)
-N_{k},  \notag
\end{eqnarray}%
this expression is not defined in the value of (EoS) $\omega =1/3$.
According to \cite{G2}, a numerical values can be obtain : $M_{p}=\kappa
^{-1}=2.435\times 10^{18}Gev,~a_{0}=1,~T_{0}=2.725K,~\bar{g}_{\ast }\simeq
106.75,~k=0.05Mpc^{-1}$, which reduces Eq. (\ref{i}) to

\begin{eqnarray}
N_{pre} &=&\left[ 60.0085-\frac{1}{4}\ln \left( \frac{3\lambda _{end}}{%
100\gamma \pi ^{2}}\right) -\frac{1}{4}\ln \left( \frac{V_{end}}{H_{k}^{4}}%
\right) -N_{k}\right]  \notag \\
&&-\frac{1-3\omega }{4}N_{th},  \label{pr}
\end{eqnarray}%
with $N_{th},$\ the reheating duration can be obtained from Eqs. (\ref{g})
and (\ref{h})

\begin{equation}
N_{th}=\frac{1}{3(1+\omega )}\ln \left( \frac{\lambda _{end}V_{end}}{\frac{%
\gamma \pi ^{2}}{30}\bar{g}_{\ast }T_{th}^{4}}\right) .  \label{Nr}
\end{equation}%
The duration of preheating$\ N_{pre}$ is linked to the inflationary
quantities through $\lambda _{end}$, $V_{end}$, $N_{k}$, and $H_{k}$. These
quantities need to be calculated for the model we considered previously in
this work. In addition to that, the preheating duration is also described by
a parameter $\gamma $ we defined previously, that connects the energy
density at the end of inflation $\rho _{end}$ to the preheating energy
density $\rho _{pre}$. $N_{th}$\ can be calculated considering the final
reheating thermalization temperature as \cite{Gb6} $T_{th}>10^{12}GeV.$
Inflation ended when the value of $\omega $ became larger than $-1/3$, in
order to satisfy the condition of density energy dominance and preserve the
causality $\omega $ must be smaller than $1$, when reheating is finished the
(EoS) reached $1/3$, for this reason, we will test if the choice of specific
values of (EoS) parameter has effects on preheating duration.

Inflation ends when the slow-roll parameters $\epsilon ,\delta _{1}$ become
as $\epsilon (\phi _{end})=1,\delta _{1}(\phi _{end})=1$. One can calculate $%
V_{end}$ and $\lambda _{end}$ using Eqs. (\ref{a}-\ref{d}), and (\ref{j})
which gives:

\begin{eqnarray}
V_{end} &=&\frac{V_{0}}{\kappa ^{4}}\left[ \frac{n^{2}}{2}(1-\alpha )\right]
^{\frac{n}{2}}, \\
~\lambda _{end} &=&-\frac{3n}{4\alpha (n+1)-2n}.
\end{eqnarray}%
The Hubble parameter at the time of horizon from the slow-roll
approximations $3H_{k}^{2}\approx \kappa ^{2}V(\phi _{k}),$ is obtained by
calculating $\phi _{k}(N_{k})$ from Eq. (\ref{N}) taking into account the
large field inflation case $(\phi _{k}\gg \phi _{end})$

\begin{equation}
\kappa \phi _{k}=\sqrt{\frac{n}{2}(1-\alpha )(4N_{k}+n)},
\end{equation}%
as a result

\begin{equation}
H_{k}^{4}=\left( \frac{V_{0}}{3\kappa ^{2}}\right) ^{2}\left[ \frac{n}{2}%
(1-\alpha )(4N_{k}+n)\right] ^{n}.
\end{equation}

It can be seen from previous results that $N_{k}$, $V_{end}$, $\lambda
_{end} $ and $H_{k}$ are all expressed in terms of spectral index $n_{s}$, $%
\alpha , $ and $n.$\ Hence $N_{pre}$ can be obtained as a function of $n_{s}$
from Eq. (\ref{pr}). 
\begin{figure*}[]
\resizebox{1\textwidth}{!}{  \includegraphics{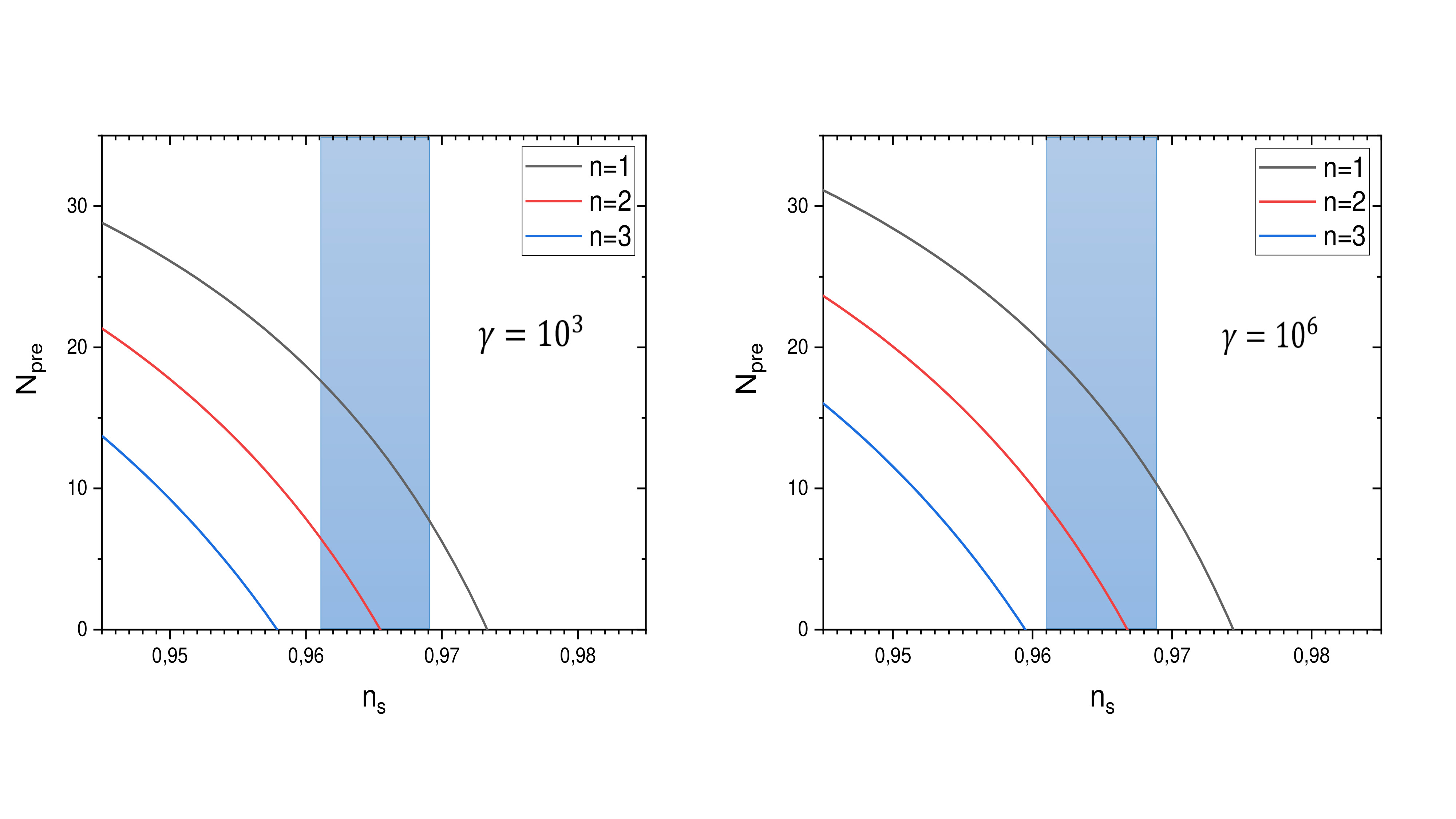}
}
\caption{The figure shows $N_{pre}$ the length of preheating, as a function
of $n_{s}$ choosing the values of $\protect\gamma $ to be $\protect\gamma %
=10^{3}$ and $\protect\gamma =10^{6}$ for power-law potential with an
inverse monomial coupling. Here, the vertical light blue region represents
Planck's bounds on $n_{s}$ \protect\cite{G2}. The black line corresponds to $%
n=1$, the red line corresponds to $n=2$, and the blue line corresponds to $%
n=3$.}
\label{fig:2}
\end{figure*}

Fig. \ref{fig:2} show the variation of the e-folds number during preheating
as a function of spectral index $n_{s}.$ We choose the three values $n=1,2$\
and $3.$\ Each curve fall at a point that corresponds to an instantaneous
preheating $(N_{pre}\rightarrow 0)$, we should mention here\ that the
preheating duration is independent of the choice of the (EoS) value $\omega $%
. As depicted in the Figure, for $\gamma =10^{3},10^{6}$, the case $n=3$\
completely lies outside the Planck bounds on $n_{s}=0.9649\pm 0.0042$ \cite%
{G2}, in order to satisfy observations, $n$\ must be bounded as $n<3.$

\section{Gravitational Waves From Preheating}

\label{sec:4} Since we are interested in the correlation of gravity-wave
energy density spectrum with current observations, We must translate the
previous GW spectrum into current physical quantities. The present scale
factor in comparison to the one when GW production stops can be expressed as 
\cite{G3,Ga4}

\begin{equation}
\frac{a_{end}}{a_{0}}=\frac{a_{end}}{a_{pre}}\left( \frac{a_{pre}}{a_{th}}%
\right) ^{1-\frac{3}{4}\left( 1+\omega \right) }\left( \frac{\bar{g}_{\ast }%
}{\bar{g}_{0}}\right) ^{-1/12}\left( \frac{\rho _{r,0}}{\rho _{\ast }}%
\right) ^{1/4}.  \label{ca}
\end{equation}

Supposing that GW production stops at the end of preheating, $pre$
represents the time when GW production is finished, "$0"$ and "$th"$
represent the present and the time when thermal equilibrium is reached,
respectively. While $\rho _{r,0}$ is\ the present radiation energy density
and the total energy density of the scalar field is represented by $\rho
_{\ast }.$ We define $\bar{g}_{\ast }/\bar{g}_{0}\simeq 106.75/3.36\simeq
31. $\ $\omega $ is the equation of state which in the Ref. \cite{Gg1} it
has been shown that $\omega $ reaches $1/3$ just after preheating, that
means that $\left( a_{pre}/a_{th}\right) ^{1-3/4\left( 1+\omega \right) }=1$%
\ since $\omega =1/3.$\ From Eq. (\ref{ca}), the corresponding physical
frequency today is given by

\begin{equation}
f=\frac{k}{2\pi a_{0}}=\frac{k_{0}}{a_{end}\rho _{\ast }^{1/4}}\times \left(
4\times 10^{10}Hz\right) ,
\end{equation}%
let us denote $k_{0}=k/a_{0}.$\ Knowing that the abundance of radiation
today given as $\Omega _{r,0}h^{2}=h^{2}\rho _{r,0}/\rho _{c,0},$\ with $h,$%
\ is the present dimensionless Hubble constant and $\Omega
_{gw,0}h^{2}\propto 1/a_{0}^{4}$ \cite{G3}. Using Eq. (\ref{ca}) and Because
GW decays like radiation with cosmic expansion, one can calculate the
present GW spectra \cite{Ga4}

\begin{equation}
\Omega _{gw,0}h^{2}=\Omega _{gw}(f)~\left( \frac{a_{end}}{a_{pre}}\right)
^{4}~\left( \frac{\bar{g}_{\ast }}{\bar{g}_{0}}\right) ^{-1/3}\Omega
_{r,0}h^{2}.
\end{equation}

The number of e-folds between the end of inflation to the time when
preheating completed can be written as \ 

\begin{equation}
\frac{a_{end}}{a_{pre}}=e^{-Npre},
\end{equation}%
to obtain the final form of gravity-wave energy density spectrum, given as
follows

\begin{equation}
\Omega _{gw}(f)=\frac{\Omega _{gw,0}h^{2}}{\Omega _{r,0}h^{2}}\left( \frac{%
\bar{g}_{\ast }}{\bar{g}_{0}}\right) ^{1/3}e^{4Npre}.
\end{equation}

\begin{figure*}[]
\resizebox{0.8\textwidth}{!}{  \includegraphics{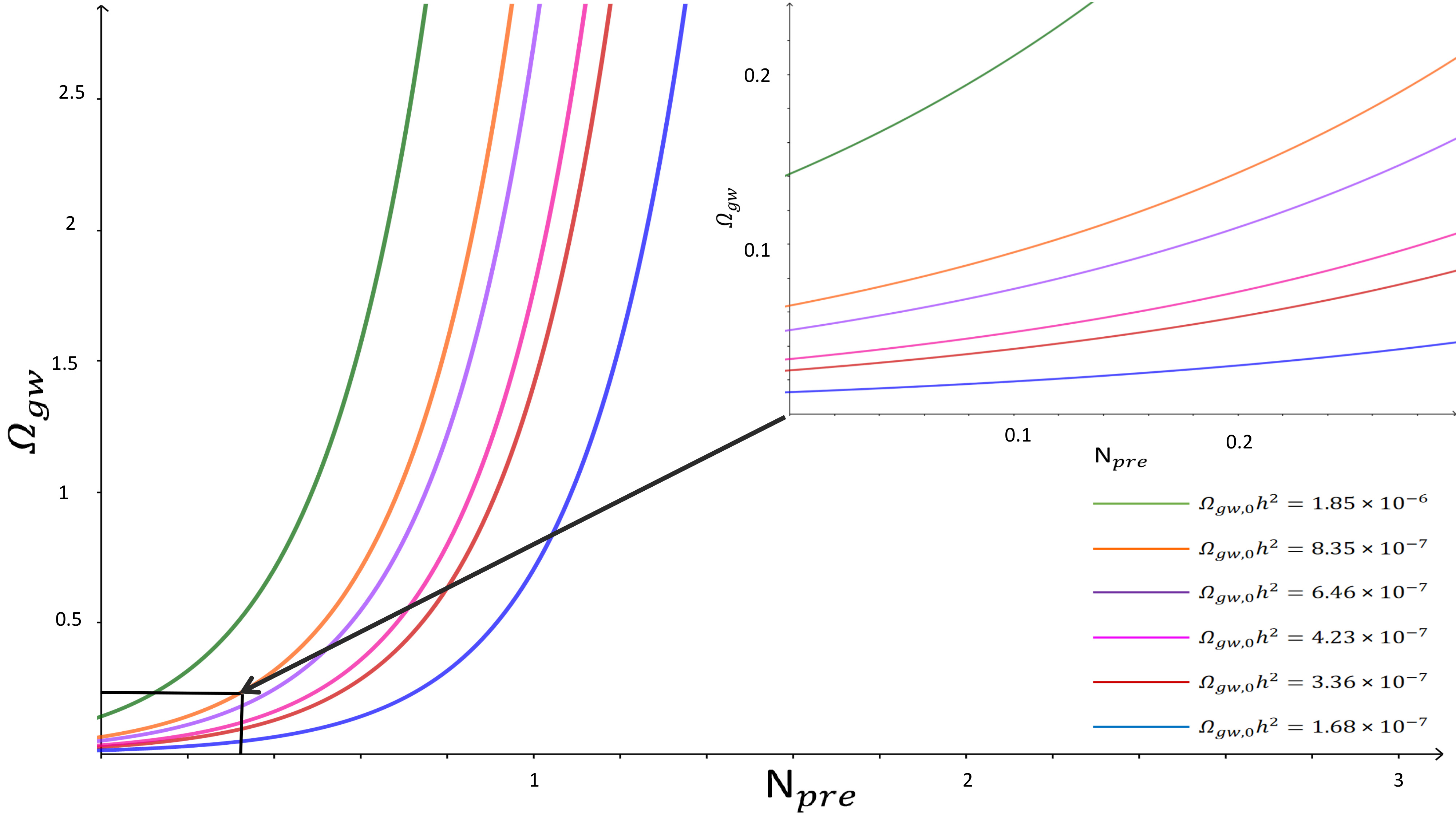}
}
\caption{The variation of the density spectra of GWs as a function of $%
N_{pre}$. For different values of gravity wave energy density $\Omega
_{gw,0}h^{2}$.}
\label{fig:3}
\end{figure*}

From Fig. \ref{fig:3} the variation of $\Omega _{gw}$\ as a function of $%
Npre $\ for some fixed values of $\Omega _{gw,0}h^{2}$\ are presented, we
plotted the energy density spectrum\ $\Omega _{gw}$\ as a function of the
preheating duration $N_{pre},$\ taking the present GW spectra to be\ $%
3.36\times 10^{-7}\leq \Omega _{gw,0}h^{2}\leq 1.85$\ $\times 10^{-6}$, when 
$N_{pre}\rightarrow 0$\ the GW energy density takes an initial value for all
the cases with different $\Omega _{gw,0}h^{2}.$ When we increase the present
GW spectra, the initial values that correspond to $\Omega _{gw}(N_{pre}=0)$
increases as well.

\begin{figure*}[]
\resizebox{1\textwidth}{!}{  \includegraphics{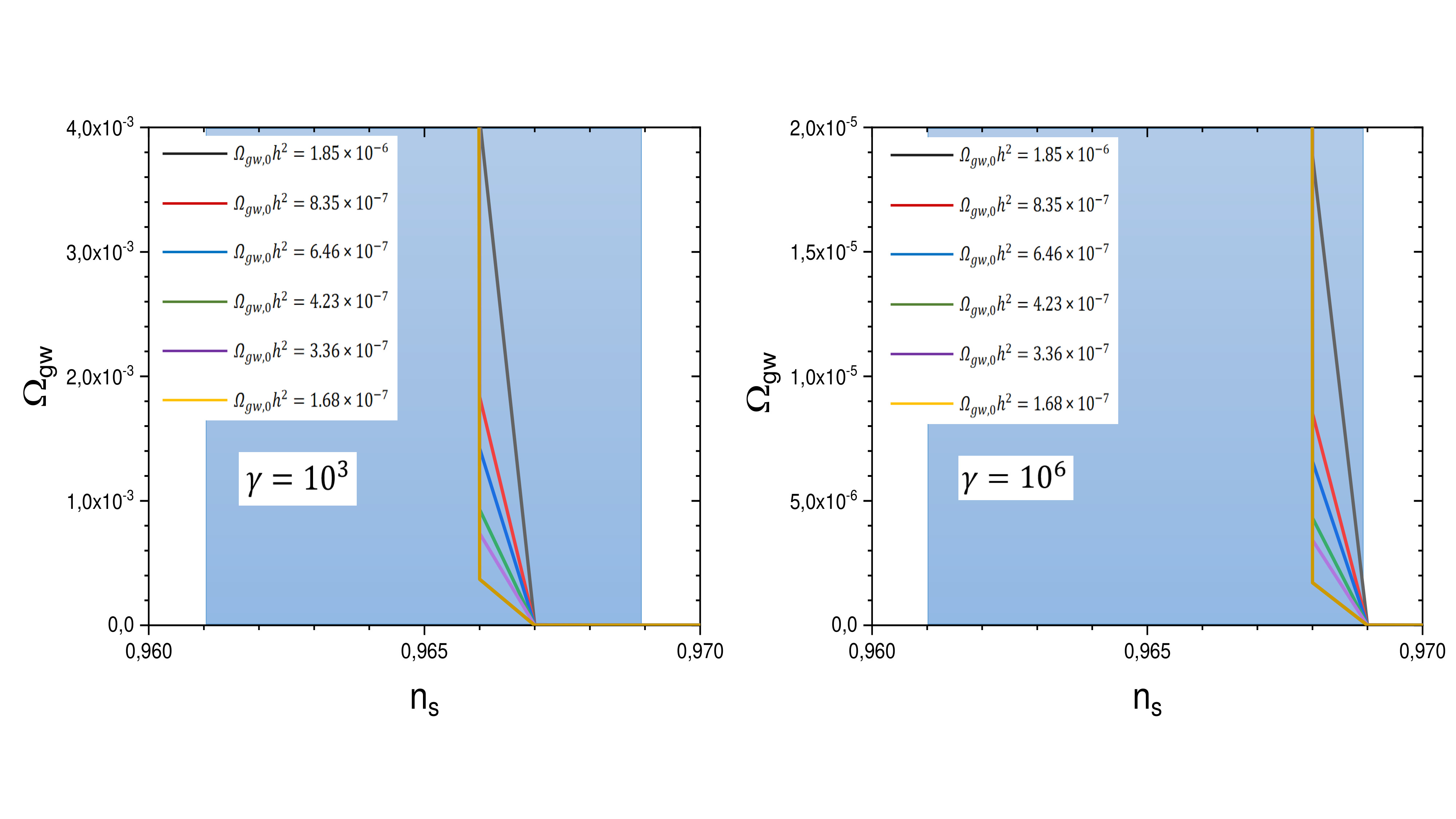}
}
\caption{The variation of the density spectra of GWs as a function of $n_{s}$%
. For different values of gravity wave energy density $\Omega _{gw,0}h^{2}$,
we choose $3.36\times 10^{-7}\leq \Omega _{gw,0}h^{2}\leq 1.85$\ $\times
10^{-6}$. The black line corresponds to $\Omega _{gw,0}h^{2}=1.85$\ $\times
10^{-6}$, red line corresponds to $\Omega _{gw,0}h^{2}=8.35$\ $\times
10^{-7} $, blue line corresponds to $\Omega _{gw,0}h^{2}=6.46$\ $\times
10^{-7}$, green line corresponds to $\Omega _{gw,0}h^{2}=4.23$\ $\times
10^{-7},$ purple line corresponds to $\Omega _{gw,0}h^{2}=3.36$\ $\times
10^{-7},$ and yellow line corresponds to $\Omega _{gw,0}h^{2}=1.68$\ $\times
10^{-7}$.}
\label{fig:4}
\end{figure*}

The variation of the GW density spectrum with respect to the spectral index $%
n_{s}$ is shown in Fig. \ref{fig:4}. Considering different values of the
current GW spectra $\Omega _{gw,0}h^{2}$, we plot $\Omega _{gw}$ considering
the expansion of the universe from the end of inflation up to later times of
preheating, because of the higher final temperature of reheating $%
T_{th}>10^{12}GeV,$ the duration $N_{th}$ from Eq. (\ref{Nr}) could be
considered as instantaneous, which make preheating duration minimally
dependent on the (EoS) parameter $\omega $ as observed in Eq. (\ref{pr}). We
choose the most compatible case from the previous analysis\ which favors the
case $n=2$ and consider the two values of $\gamma :$\ $10^{3},\ $and\ $%
10^{6}.$\ It's easy to see that the density spectrum curves with both values
of $\gamma \ $are compatible with observations according to Planck's
results, the curves decrease away from the observation bound when the GW
energy density became very negligible $\Omega _{gw}\rightarrow 0$. For the
case $\gamma =10^{3},$ with $\Omega _{gw}\geq 4\times 10^{-3}$ all the lines
with different $\Omega _{gw,0}h^{2}$\ tends towards the value of spectral
index $n_{s}=0.966$. However, the case where $\gamma =10^{6}$ the curves
converge to the value $n_{s}=0.968$\ when $\Omega _{gw}\geq 1.8\times
10^{-5} $.

\section{Conclusion}

\label{sec:5} After we review the basic equations that describe GB
inflation, we discuss the power-law model of inflation with inverse monomial
GB coupling. The expression of the observational parameters $n_{s},$ and $r$
were calculated, we computed these parameters as functions of inflation $e$%
-folds $N_{k}$ for the power-law potential with an inverse monomial model.
We review the basics of Primordial GWs, then the energy density carried by
these waves was calculated. We derived the preheating duration as functions
of inflationary Gauss-Bonnet parameters in Eq. (\ref{pr}), and consider the
thermalization temperature as $T_{th}>10^{12}GeV$. We numerically estimated
the preheating parameters using our analytic results. Knowing that it's
independent of the choice of the (EoS), the duration of preheating is
plotted as a function of the spectral index for the model we considered
previously, and showed that it's sensitive to the parameters $\gamma $\ and $%
n$. We finally calculated the gravity-wave energy density spectrum as a
function of the durations $N_{pre},$\ which is a\ possible way to study GW
density spectrum according to recent Planck's results.\ Assuming the density
parameter $\Omega _{gw,0}h^{2}$ to be $3.36\times 10^{-7}\leq \Omega
_{gw,0}h^{2}\leq 1.85$\ $\times 10^{-6},$ we chose $n=2$ and $\gamma
:~10^{3},~10^{6},$ and\ found that both cases where $\gamma =10^{3},10^{6}$
show good consistency with observation. We conclude that the GB term appears
to be important not only during inflation but also during later phases such
as preheating, regardless of whether the process is instant or takes a
certain number of e-folds to complete, once we determine the final
temperature of thermalization $T_{th}$, other preheating parameters are
determined using a variety of inflation models. As a result, it would be
interesting to investigate the physics of preheating in the context of PGW.

\end{document}